\documentstyle[11pt,IAUS212,twoside,epsf]{article}
\begin{document}

\markboth{Strickland, Heckman, Colbert, 
Hoopes \& Weaver}{The hot and warm phases of the ISM}
\pagestyle{myheadings}
\setcounter{page}{1}

\title{Recent progress in understanding the hot and warm gas phases
in the halos of star-forming galaxies}
\author{D.K. Strickland\altaffilmark{1}, T.M. Heckman, E.J.M. Colbert,
C.G. Hoopes and}
\affil{Dept. of Physics \& Astronomy,
The Johns Hopkins University, 3400 N. Charles St., Baltimore, MD 21218, USA.}
\altaffiltext{1}{{\it Chandra} Fellow}

\author{K.A. Weaver}
\affil{NASA/GSFC, Code 662, Greenbelt, MD 20771, USA.}

\begin{abstract}

In this contribution we present a few selected examples of how the
latest generation of space-based instrumentation -- 
NASA's {\it Chandra} X-ray Observatory and 
the Far-Ultraviolet Spectroscopic
Explorer ({\it FUSE}) -- are finally answering old questions about the
influence of massive star
feedback on the warm and hot phases of the ISM and IGM.
In particular, we
discuss the physical origin of the soft thermal X-ray emission in the
halos of star-forming and starburst galaxies, its relationship to
extra-planar H$\alpha$ emission, and plasma diagnostics using
{\it FUSE} observations of O {\sc vi} absorption and emission.
\end{abstract}

\section{Introduction -- the hot phases as viewed by Chandra and FUSE}

Massive stars exercise a profound influence over the baryonic component
of the Universe, through their return of ionizing radiation, and
via supernovae (SNe), kinetic energy and metal-enriched gas, back into the
ISM from which they form --- usually called ``feedback''. 
Feedback influences gas-phase conditions in the immediate
environment of the clusters within which the massive stars form, 
on galactic-scales the phase structure and energetics of the ISM,
and on multi-Mpc scales the thermodynamics and enrichment of
the inter-galactic medium (IGM).

The vast range of spatial scales involved is only one of the difficulties 
encountered in attempting to study feedback. Another is the broad range of
complicated gas-phase physics -- (magneto)hydrodynamic
effects such as shocks and turbulence, thermal conduction, and non-ionization
equilibrium emission processes. 
A final complication is that much of the energy and
metal-enriched material involved is in the hard-to-observe coronal 
($T \ga 10^{5}$ K) and hot ($T \ga 10^{6}$ K) gas phases.

\begin{figure}[!t]
\plotone{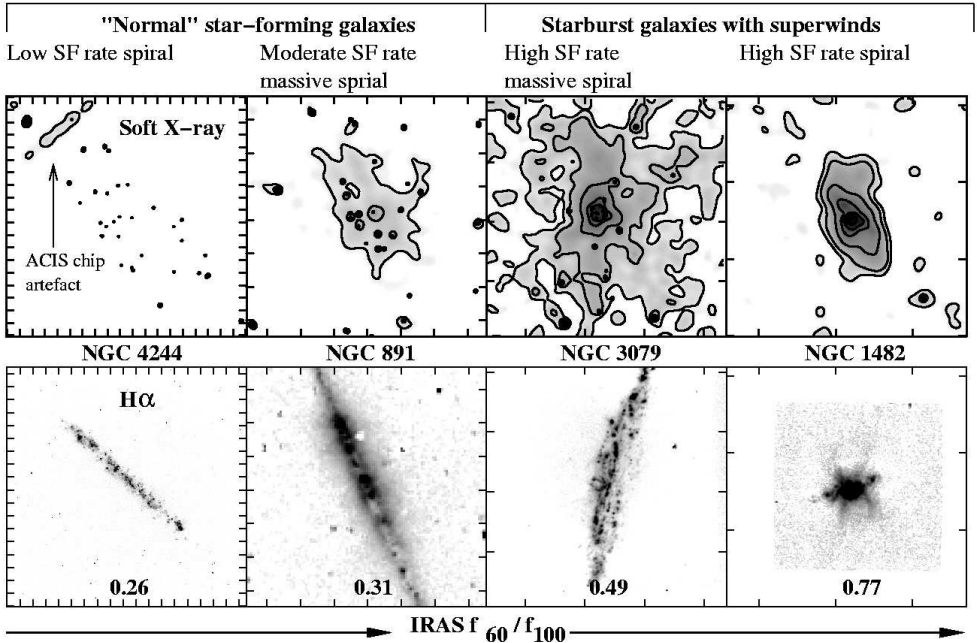}
\caption[Sample galaxies]{A few examples from our larger survey
	of edge-on star-forming galaxies (Strickland et al 2002b).
	The top panels plot {\it Chandra} ACIS-S
	0.3 -- 2.0 keV X-ray surface brightness. All galaxies are shown
	on the same logarithmic intensity scale, with contours spaced by
	0.5 dex. Point sources have been left in these images, but are removed
	for scientific analysis.
	The lower panels are continuum-subtracted 
	H$\alpha$+[N {\sc ii}] emission, with the IRAS $f_{60}/f_{100}$ ratio
	(a measure of the SF intensity) given
	at the bottom of the image. Each image shows a $20 \times 20$ 
	kpc region. Tick-marks 
	represent $1\arcmin$ --- {\em Chandra}'s resolution
	is sub-arcsecond.
	}
\label{fig:dks:sample_xray_ha}
\end{figure}

\section{X-ray emission mechanisms and the micro-physics of superwinds}
\label{sec:micro}

Local starburst galaxies show unambiguous evidence for multi-phase,
$\sim 500$ km/s, bipolar outflows, characterized by $\ga 10$ kpc-scale
extra-planar soft thermal X-ray, optical recombination and non-thermal
radio emission (see Heckman, Armus \& Miley 1990). These superwinds are driven by the pressure of the thermalized
ejecta from the large numbers of core-collapse supernovae resulting
for the starburst event. Superwinds, as perhaps the most-extreme, and
least subtle, form of feedback are an ideal place to explore and
test our understanding of the
many physical processes encompassed within the subject. 
The relationship between superwinds and the extra-planar diffuse ionized gas
or eDIG (Rand 1998; Dettmar 1993) 
in normal spiral galaxies, where kinematic evidence 
for outflow is lacking, is currently unclear.

\begin{figure}[!t]
\plotone{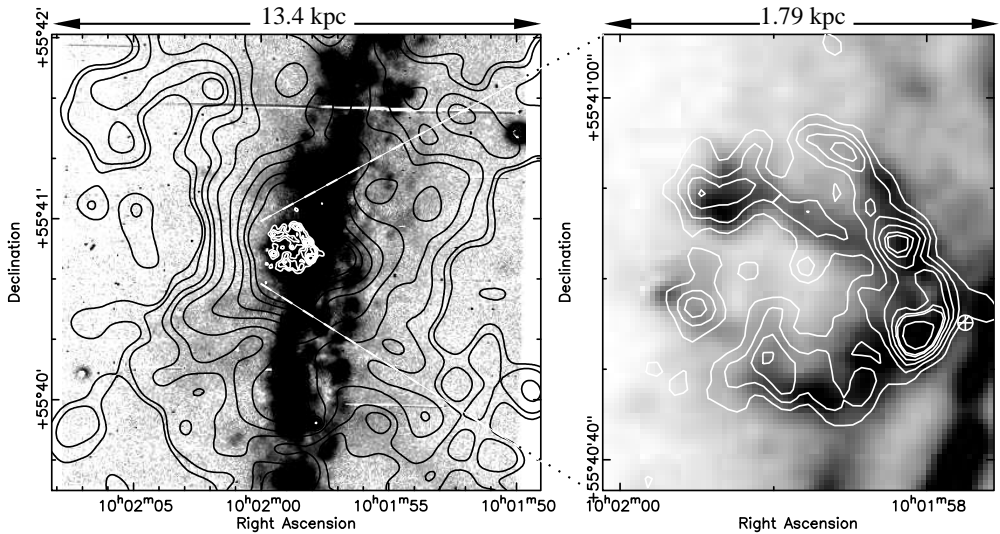}
\caption[NGC 3079]{The diffuse X-ray (contours) and H$\alpha$ (grey-scale)
	emission in the starburst/LINER galaxy NGC 3079 correlate 
	extremely well, in common with the other starbursts in our sample.
	The X-ray emission is smoothed 
	0.3--2.0 keV diffuse emission from our Chandra observation (all
	point sources have been removed). Left-hand-panel: X-ray features
	are locally significant with S/N=3. The H$\alpha$+[N {\sc ii}] image,
	showing the large-scale filaments, is from Lehnert \& Heckman (1996).
	Right-hand panel: The
	nuclear filaments, with both X-ray and H$\alpha$ images (recent
	APO 3.5m data) smoothed
	to a final effective resolution of $1\farcs4$ ($\approx 116$ pc).
	The location of the LLAGN is marked by the open circle
	with a central cross.}
\label{fig:dks:n3079}
\end{figure}
 
A quantitative assessment of the energetics, mass and composition
of superwinds, and possibly galactic fountains in non-starburst
galaxies, depends on understanding the physical origin of the thermal
X-ray emission in winds. Prior to {\it Chandra}, the poor spatial
resolution of X-ray telescopes made it impossible to determine if the
X-ray emission came directly from the (possibly mass-loaded, see below)
volume-filling wind of SN ejecta (Suchkov et al. 1996), or
merely from low-filling factor ambient disk or halo gas over-run by the
wind (Suchkov et al. 1994; Strickland \& Stevens 2000). 
Low spatial resolution also complicated X-ray spectral analysis, due
contamination by unresolved point sources and the blending together
of physically distinct regions (Weaver, Heckman \& Dahlem 2000).

\subsection{Spatially-correlated X-ray/H$\alpha$ emission}

Within the central kiloparsec
of NGC 253, NGC 4945, NGC 3079 (see Fig.~\ref{fig:dks:n3079})
and M82 (Strickland et al 2000; Strickland 2001;
Schurch, Roberts \& Warwick 2002; Strickland et al. 2002b)
{\it Chandra} observations demonstrate the thermal X-ray emission
matches up almost exactly to the filamentary, limb-brightened 
H$\alpha$ emission. This proves that the X-ray emission arises in
some form of interaction between the currently-invisible wind and
the denser, cooler, ambient ISM responsible for the H$\alpha$ emission.

\begin{figure}[!t]
\plotone{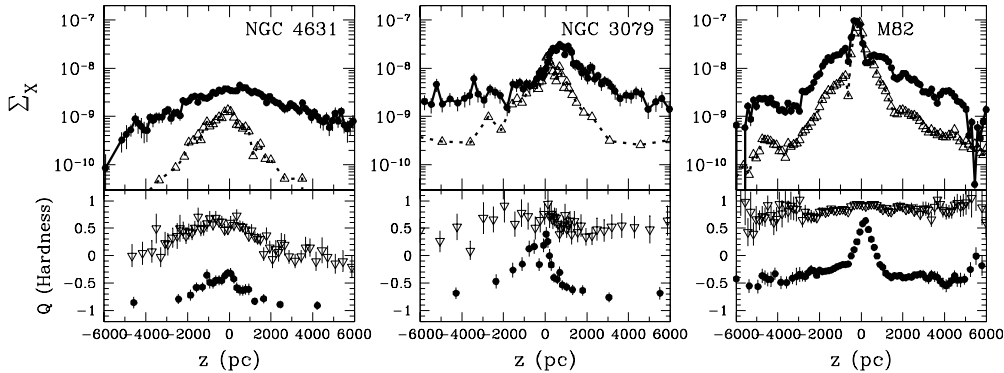}
\caption[SF intensity]{Top panels: Minor axis diffuse X-ray surface 
brightness profiles (circles: 0.3--1.0 keV energy band, 
triangles: 1.0--2.0 keV), for three of galaxies in our survey 
(Strickland et al. 2002b). 
{\em Emission from point sources, 
and the X-ray background, has been removed}. 
The diffuse emission is best
characterized as an exponential with a scale height of several kpc.
Power law surface brightness distributions are statistically unacceptable.
Galaxies are ordered from left to right in terms of increasing
IRAS $f_{60}/f_{100}$ ratio.
Lower panels: Spectral hardness ratios $Q_{A}$ (triangles) and $Q_{B}$ 
(circles). See Strickland
et al (2002a) for a published definition of these ratios and their
significance. Note the spectral uniformity outside the central disk
region.}
\label{fig:dks:zprof}
\end{figure}

On larger, $\sim 10$ kpc scales, we have also convincing evidence that the
X-ray emission is associated with H$\alpha$ emission. The close association 
between {\em halo} X-ray and H$\alpha$ emission has been know about
in M82 for many years (see Dahlem, Weaver \& Heckman 1998, and references
therein), but we have also found X-ray emission associated
with faint H$\alpha$ filaments in the halos of NGC 253 
(Strickland et al. 2002a), 
NGC 3628 (Strickland 2001) and NGC 3079. Any physical model for superwinds,
in particular numerical models of outflows, must explain and reproduce this
X-ray/H$\alpha$ spatial correlation. Strickland et al. (2002a) discuss
a variety of models that satisfy this criterion. On such model is that
the H$\alpha$ emission arises in a radiative shock driven into a
several kpc-scale-height thick disk or halo medium, the X-ray emission
coming from a reverse shock driven into the superwind. The 
semi-exponential surface brightness, and isothermal nature, of the
minor axis X-ray emission we observe in many superwinds 
(see Fig.~\ref{fig:dks:zprof}) is
consistent with this model.

\subsection{Shocks and the superwind X-ray/H$\alpha$ flux ratio}

Not only is there a strong similarity in morphology between the
soft X-ray and optical H$\alpha$~emission in superwinds,
but the X-ray-to-H$\alpha$ flux ratio always appears to between
within a factor $\sim2$ of unity throughout any
superwind. Strickland et al. (2002a) discusses this with respect to
NGC 253, the northern ``cap'' in M82 (Lehnert, Heckman \& Weaver 1999) 
and Arp 220. We have,
as reported at several prior conferences, also
found $f_{\rm X}/f_{\rm H\alpha} \sim 1$ in NGC 3079's nuclear bubble,
and generally throughout the NGC 1482 and M82 winds. Note that
as the extinction $A_{\rm H\alpha} \approx A_{\rm 0.8 keV}$, spatial
variations in the extinction over the superwind do not alter 
$f_{\rm X}/f_{\rm H\alpha}$ significantly.

\begin{figure}[!t]
\plotone{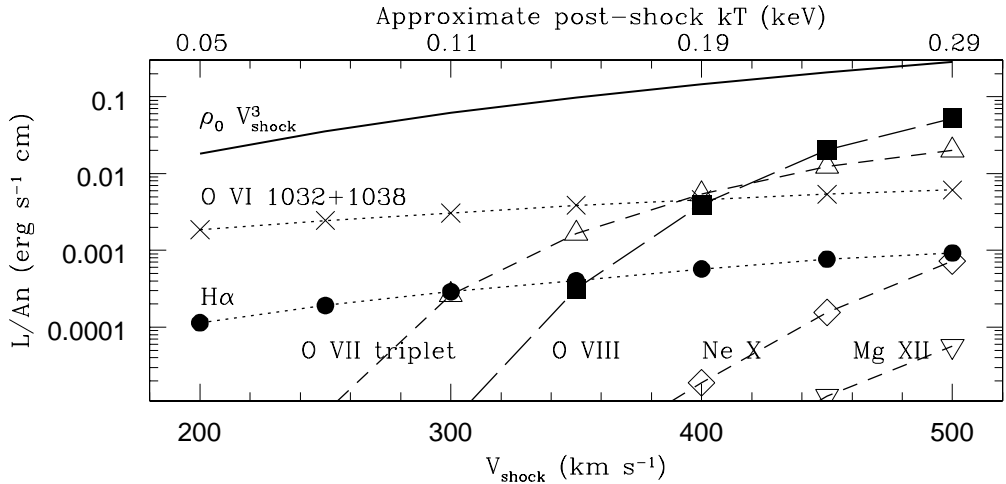}
\caption[Shock model]{Density-normalized flux per unit area for a
	{\em fully radiative} 
	high velocity shock, based on the Dopita \& Sutherland
	(1996) models. Emission from\hbox{ O {\sc vii}} and O {\sc viii} 
	provide a large fraction of the total X-ray emission within
	the range of post-shock temperatures displayed. This simple
	model fails to explain X-ray \& H$\alpha$ emission in superwinds.}
\label{fig:dks:shock}
\end{figure}

This result, in particular that that it appears to be a general rule over
a variety of spatial-scales 
in many, possibly all, superwinds, is very significant. 
It can be used constrain
the physical mechanisms responsible for the soft X-ray (and to a 
certain extent, the optical) emission in superwinds. For example,
we can rule out a model where the tenuous superwind passes through a
fully-radiative internal
high-velocity shock, compressing and heating it to X-ray-emitting 
temperatures, followed by downstream cooling to $T \sim 10^{4}$ K 
and H$\alpha$ emission. In such a model $f_{\rm X}/f_{\rm H\alpha} \approx 1$
{\em only} for $v_{\rm shock} \sim 260$ -- 300 km/s
(see Fig.~\ref{fig:dks:shock}), 
{\em but} this predicts X-ray temperatures
$kT \approx 0.1$ keV, well below the broad temperature range observed
of $kT \sim 0.2$ -- 0.7 keV. 
The predicted ratio of O {\sc vii} 
to O {\sc viii} emission is also an order of magnitude greater than 
that observed (see Strickland et al. 2002a). Other shock-related
models (e.g. Lehnert et al. 1999) are
quantitatively more successful (Strickland et al. 2002c).

\subsection{X-ray-emission from mass-loaded winds revisited}

One of the original aims of our Chandra observations of local superwinds
was to see if we could spatially separate the volume-filling wind fluid
from X-ray emission due to wind/ISM interactions. Only in the central
few hundred parsecs might the wind fluid be bright enough to detect,
necessitating arcsecond spatial resolution X-ray spectral-imaging.

In NGC 253, the closest bright starburst, we placed an upper limit of
20\% of the X-ray emission coming from any volume-filling component 
(Strickland et al 2000), the overwhelming majority of the observed emission
coming from a spectrally-uniform limb-brightened conical structure.
Recently, Pietsch et al (2001) and Schurch et al. (2002)
claim that the degree of limb-brightening in NGC 253 and NGC 4945
decreases at lower X-ray energies.
Unfortunately these authors did not quantify the effect,
but they interpret this as due to the increasing importance of
emission from a ``mass-loaded'' wind\footnote{Mass-loaded is 
used here in the sense employed by Suchkov et al. (1996), where the
X-ray emission from a superwind was modelled as coming from
a volume-filling fluid, the density of which had been 
increased by {\em uniformly} mixing in material from the ISM the wind had
over-run.} component at lower X-ray energies.

We briefly mention another plausible interpretation for this
observed effect, which we defend quantitatively in Strickland et al. (2002d).
The column density of material is largest when looking at the
limb of these outflow cones. Optical and FIR observations 
(e.g. Phillips 1993; Radovich, Kahanp\"a\"a \& Lemke (2001)) show
that the dense filamentary material in the walls of superwinds contains
significant amounts of neutral gas and dust. 
Soft X-rays coming from the limbs of the outflow
cones will thus suffer more absorption than those coming from the front
and rear walls. At higher X-ray energies this differential
absorption is reduced, the net effect being to reduce the apparent
degree of limb-brightening at lower X-ray energies {\em without}  
mass-loading. 

\section{Gas in the coronal phase and O {\sc vi}}

Radiative energies losses, in any realistic multi-phase model for the ISM,
are dominated by gas at temperatures between $10^{5}$ and $10^{6}$ K,
even if the majority of the energy is stored in hotter gas. A particularly
powerful probe of gas in this ``coronal'' phase is the O {\sc vi} doublet
at $\lambda = 1032$ and $1038 \AA$. Approximately 20\% of the total cooling
at $T \sim 3\times10^{5}$ K (in a Solar abundance plasma)
is due to this doublet alone.
With the high sensitivity and spectral resolution of {\it FUSE}, we can,
for the first time, probe superwind kinematics in the hot phases, 
and directly see metals in outflow.

\subsection{O{\sc vi} in absorption}

In the proto-typical starbursting dwarf galaxy NGC 1705, {\it FUSE}
observations (Heckman et al 2000) 
show that the coronal phase gas is flowing out of the galaxy
at higher velocity than the warm ionized gas at $T \sim 10^{4}$ K, which
in turn are flowing outwards at higher velocity than the warm neutral
medium. That the hot phases can have higher outflow velocities is of
great physical significance, as it supports theoretical
claims that the hot, metal-enriched, phases can be preferentially ejected
into the IGM.

Heckman et al. (2002) demonstrate that the O {\sc vi} column density,
in a radiatively cooling plasma,
is largely independent of the number density and metallicity of the gas,
and depends only on a characteristic
velocity in the cooling gas. 
Distinct physical processes, such as thermally-unstable
collapsing gas clouds, 
gas in radiative shocks, turbulent mixing layers and
conductive interfaces {\em all produce the essentially the 
same O {\sc vi} column density}! This results applies in general to
many high-ionization lines, 
and complicates the use of the
column densities {\em alone} as a diagnostic tool. For example
the presence and column densities of high ionization
species in wind-blown bubbles (Boroson et al. 1997) {\em do not}
provide unambiguous evidence for presence and
action of conductive interfaces.

\subsection{O{\sc vi} in emission}

Emission lines from coronal phase gas can be used to place strong constraints
on various physical models for the X-ray \& optical emission from
superwinds. 
For example, the $\sim 300$ km/s radiative shock 
model
(Fig.~\ref{fig:dks:shock}) predicts an O {\sc vi}
line luminosity approximately an order of magnitude more luminous than
the total soft X-ray emission, or the optical H$\alpha$ emission.

The application of such diagnostics to superwinds has only just become
possible with the launch of FUSE. We (PI: Hoopes) have begun a 
program looking for
O {\sc vi} emission at various locations in M82 and NGC 253's winds, where
we know the optical and X-ray properties from our existing data.

\section{Summary}

In the three years since their launch in mid-1999, the un-matched
capabilities of both {\it Chandra} and {\it FUSE}, have allowed
subtantial progress to be made in understanding the physics of the
feedback in star-forming galaxies. Old questions
have been answered, and new ones have arisen. 
With several, no-doubt productive,
years left for each instrument, it is likely that progress will continue
to be rapid in 
understanding the relationship between massive stars and the 
warm and hot phases of the ISM and IGM.

\acknowledgements
We would like to thank S.~Hameed for the use of his H$\alpha$
image of NGC 1482. DKS is supported by NASA through 
{\it Chandra} Postdoctoral Fellowship Award
Number PF0-10012, issued by the {\it Chandra} X-ray Observatory Center.

\end{document}